\documentclass[aip,jmp,showpacs,nofootinbib]{revtex4}
\usepackage{latexsym}
\usepackage{amsfonts}
\usepackage{amsbsy}
\usepackage{mathrsfs}

\begin{document}

\title{Topological field theories in $n$-dimensional spacetimes and
Cartan's equations}

\date{\today}

\author{Vladimir Cuesta$^1$}
\email{vladimir.cuesta@nucleares.unam.mx}

\author{Merced Montesinos$^2$}
\email{merced@fis.cinvestav.mx}

\author{Mercedes Vel\'azquez$^2$}
\email{mquesada@fis.cinvestav.mx}

\author{Jos\'e David Vergara$^1$}
\email{vergara@nucleares.unam.mx}

\affiliation{$^1$Instituto de Ciencias Nucleares, Universidad
Nacional Aut\'onoma de M\'exico, 70-543, Ciudad de M\'exico, M\'exico\\
$^2$Departamento de F\'{\i}sica, Centro de
Investigaci\'on y de Estudios Avanzados del Instituto Polit\'ecnico
Nacional, Instituto Polit\'ecnico Nacional 2508, San Pedro Zacatenco, 07360, Gustavo A. Madero, Ciudad de M\'exico, M\'exico}

\begin{abstract}
Action principles of the BF type for diffeomorphism invariant topological field theories living in $n$-dimensional spacetime manifolds are presented. Their construction is inspired by Cuesta and Montesinos' recent paper where Cartan's first and second structure equations together with first and second Bianchi identities are treated as the equations of motion for a field theory. In opposition to that paper, the current approach involves also auxiliary fields and holds for arbitrary $n$-dimensional spacetimes. Dirac's canonical analysis for the actions is detailedly carried out in the generic case and it is shown that these action principles define topological field theories, as mentioned. The current formalism is a generic framework to construct geometric theories with local degrees of freedom by introducing additional constraints on the various fields involved that destroy the topological character of the original theory. The latter idea is implemented in two-dimensional spacetimes where gravity coupled to matter fields is constructed out, which has indeed local excitations.
\end{abstract}


\maketitle

\section{Introduction}\label{intro}
There is a renewed interest in the study of BF theory and general relativity written as a constrained BF theory motivated by the progress of the spin foam \cite{spinfoam} and loop quantum gravity \cite{lqg} approaches to the nonperturbative and background-independent quantization of gravity. On one hand, the topological nature and the diffeomorphism invariance of BF theory makes it a suitable laboratory to test technical as well as conceptual issues related with classical and quantum gravity. On the other, there are still several issues connecting pure BF theory and BF gravity that deserve to be explored deeply.

In this context, it was recently proposed in Ref. \cite{cuesta} that Cartan's first and second structure equations together with first and second Bianchi identities can be interpreted as equations of motion for the tetrad, the connection, and a set of two-form fields $T^I$ and $R^I\,_J$. It was shown that these equations define a topological field theory, which can be obtained from an action principle of the BF type. Moreover, four-dimensional general relativity was obtained there by doing a suitable modification of the original action principle that destroyed its topological character and at the same time allowed the degrees of freedom for gravity to arise. In this way, the results and the philosophy of the paper \cite{cuesta} is that Cartan's equations encode a topological field theory and that the topological property of the theory disappears once Einstein's equations are brought into the framework.

With this in mind, it is natural to ask if the theoretical framework developed in Ref. \cite{cuesta} can be naturally extended to arbitrary finite-dimensional spacetimes. The answer is in the affirmative, this being one of the two main results reported in this paper. In fact, inspired by those results, Cartan's equations are supplemented with auxiliary fields $\phi_I$, $\phi_{IJ}$, $\psi_I$, and $\psi_{IJ}$ in such a way that the largest set of equations of motion define diffeomorphism invariant topological field theories. The new theories living on $n$-dimensional spacetime manifolds contain pure BF theory having $SO(m)$ or $SO(m-1,1)$ structure groups as particular cases. It is important to emphasize that the auxiliary fields $\phi$'s and $\psi$'s  are not involved in the framework \cite{cuesta}, this is a major difference between these approaches.

Following the viewpoint of Ref. \cite{cuesta}, it is also natural to ask whether or not it is also possible to relate the topological theories mentioned in the previous paragraph to known or new geometric theories having a nonvanishing number of local excitations by means of the introduction of geometric or algebraic relationships among the field variables involved. Once again the answer is in the affirmative, and the idea is explicitly implemented by building up a model with local degrees of freedom in two-dimensional spacetimes starting from a topological field theory. This is the second result of this paper. It is very interesting conceptually because it opens the possibility of applying the same idea in $n$-dimensional spacetimes to try to get new formulations for gravity or to build suitable modifications of it that might be also worthwhile and interesting enough.

This paper is organized as follows: action principles for diffeomorphism invariant field theories living in $n$-dimensional spacetime manifolds are given in Sec. \ref{theory}, their canonical analyses are carried out and it is explicitly shown that the theories are topological; the two-dimensional and three-dimensional cases are reported in the Appendices \ref{aa} and \ref{ab}, respectively. In Sec. \ref{particular} it is shown that by imposing suitable restrictions on the fields it is possible to build additional topological field theories from the original action principles, and in Sec. \ref{beauty} it is shown that the theories introduced in Secs. \ref{theory} and \ref{particular} are just particular members of the largest class of diffeomorphism invariant topological field theories whose gauge group need not be the orthogonal group of the vielbeins. Finally, in Sec. \ref{local}, using the results of the previous sections, a theory with local degrees of freedom which lives on two-dimensional spacetimes and resembles two-dimensional gravity plus matter fields is constructed out. This model can also be written as two interacting BF theories. The conclusions and perspectives are collected in Sec. \ref{fin}.

\section{Action principles of the BF type}\label{theory}
In the first part of this paper topological field theories of the BF type will be studied. The main goal is to build an action that reproduces as field equations in $n$ dimensions Cartan's first and second structure equations. To that end, four different types of auxiliary fields are introduced. The first one $\phi_I$ is a $(n-2)$-form that will be associated to the basis of vielbeins $e^I$; the second one $\phi_{IJ}=-\phi_{JI}$ is also a $(n-2)$-form that will be associated to the Lorentz connection $\omega^{IJ}$. The other two auxiliary fields $\psi_I$ and $\psi_{IJ}=-\psi_{JI}$ are $(n-3)$-forms that will be associated to the ``torsion'' $T^I$ and the ``curvature'' $R^{IJ}$, respectively. Note that $R^I\,_J$ is {\it not} the same as $R^I\,_J [\omega]$: $R^{IJ}$ is a set of two-forms while $R^I\,_J [\omega]$ is the curvature of $\omega^I\,_J$ (an analog comment applies to $T^I$, see Ref. \cite{cuesta} for more details). In this way, for $n$-dimensional spacetimes ${\mathcal M}^n$ with $n\geq 3$, the field theories studied in this paper are defined by the action principle
\begin{eqnarray}\label{action}
S[\omega^I\,_J, e^I, T^I, R^I\,_J, \phi_I, \phi_{IJ}, \psi_I, \psi_{IJ}] &=& \int_{\mathcal{M}^n} \left [ \phi_I \wedge \left ( d e^I + \omega^I\,_J \wedge e^J - T^I \right ) + \phi_{IJ} \wedge \left ( d \omega^{IJ} + \omega^I\,_K \wedge \omega^{KJ} - R^{IJ} \right ) \right.\nonumber\\
&& + \psi_I  \wedge \left ( d T^I + \omega^I\,_J \wedge T^J - R^I\,_J \wedge e^J \right ) \nonumber\\
&& \left. + \psi_{IJ} \wedge \left ( d R^{IJ} + \omega^I\,_K \wedge R^{KJ} + \omega^J\,_K \wedge R^{IK} \right ) \right ],
\end{eqnarray}
where $\omega^{IJ}=- \omega^{JI}$ is a Lorentz (or Euclidean) connection valued in the $so(n-1,1)$ or $so(n)$ Lie algebra, $e^I$ is a basis of one-forms, $T^I$ is a set of $n$ two-forms, $R^{IJ}=-R^{JI}$ is a set of $n(n-1)/2$ two-forms. The indices $I,J,K,\dots, $ are raised and lowered with the Minkowski ($\sigma=-1$) or Euclidean ($\sigma=+1$) metric $(\eta_{IJ}) = \mbox{diag} (\sigma, +1,+1, \ldots, +1)$ (see Ref. \cite{jmp2006} for the canonical analysis of BF theory with structure group $SO(3,1)$, Refs. \cite{iopmm,cqg2006} for alternative action principles for $SO(3,1)$ BF theory, and Ref. \cite{cqg2003} for the study of its symmetries).

The equations of motion that follow from the variation of the action (\ref{action}) with respect to the independent fields are
\begin{eqnarray}\label{emotion}
&& \delta \phi_I:  d e^I + \omega^I\,_J \wedge e^J - T^I =0, \nonumber\\
&& \delta \phi_{IJ}: d \omega^{IJ} + \omega^I\,_K \wedge \omega^{KJ} - R^{IJ} =0, \nonumber\\
&& \delta \psi_I: d T^I + \omega^I\,_J \wedge T^J - R^I\,_J \wedge e^J=0, \nonumber\\
&& \delta \psi_{IJ}:d R^{IJ} + \omega^I\,_K \wedge R^{KJ} + \omega^J\,_K \wedge R^{IK}=0, \nonumber\\
&& \delta T^I: \phi_I + (-1)^{n-3} D \psi_I=0, \nonumber\\
&& \delta R^{IJ}: \phi_{IJ} + (-1)^{n-3} D \psi_{IJ} + \frac12 \left ( \psi_I \wedge e_J - \psi_J \wedge e_I \right ) =0, \nonumber\\
&& \delta \omega^{IJ}: (-1)^{n-2} D\phi_{IJ} + \frac12 \left ( \phi_I \wedge e_J - \phi_J \wedge e_I \right ) - \frac12 \left ( \psi_I \wedge T_J - \psi_J \wedge T_I \right ) - \left ( \psi_{IK} \wedge R_J\,^K - \psi_{JK} \wedge R_I\,^K\right ) = 0, \nonumber\\
&& \delta e^I: (-1)^{n-2} D \phi_I + \psi_J \wedge R^J\,_I =0,
\end{eqnarray}
where $D$ is the covariant derivative computed with respect to the connection $\omega^I\,_J$.

In two-dimensional spacetimes ${\mathcal M}^2$, on the other hand, only Cartan's first and second structure equations are allowed because there is no room for the first and second Bianchi identities, i.e., the terms involving the $\psi$'s  in Eq. (\ref{action}) are not allowed. Consequently, the natural action principle is given by
\begin{eqnarray}\label{accion}
S[\omega^I\,_J, e^I, T^I, R^I\,_J, \phi_I, \phi_{IJ}] &=& \int_{\mathcal{M}^2} \left [ \phi_I  \left ( d e^I + \omega^I\,_J \wedge e^J - T^I \right ) + \phi_{IJ} \left ( d \omega^{IJ} + \omega^I\,_K \wedge \omega^{KJ} - R^{IJ} \right ) \right],
\end{eqnarray}
with the corresponding interpretation for the fields involved: $\omega^I\,_J$ is an $so(2)$ or an $so(1,1)$ connection one-form, $\phi_I$ and $\phi_{IJ}$ are two and one 0-forms, respectively, etc. Note that connection $\omega^I\,_J $ involved in the action principles (\ref{action}) and (\ref{accion}) is not flat, its curvature is equal to the two-form field $R^{IJ}$ as it follows from the variation of the actions (\ref{action}) and (\ref{accion}) with respect to $\phi_{IJ}$.

In order to count the number of degrees of freedom, the canonical analyses of the theories (\ref{action}) and (\ref{accion}) are performed. Let $(x^{\mu})= (x^0,x^a)= (x^0,x^1,\ldots, x^{n-1})$ be local coordinates on $\mathcal{M}^n$, which is assumed to be of the form $\mathcal{M}^n= \mathcal{S} \times \mathbb{R}$; the coordinate time $x^0$ labels the points along $\mathbb{R}$ and the space coordinates $x^a$ label the points on $\mathcal{S}$, which is assumed to have the topology of $S^{n-1}$. The canonical analyses of the field theories on ${\mathcal M}^2$ and ${\mathcal M}^3$ are explicitly carried out in the Appendices \ref{aa} and \ref{ab}, respectively. Even though the case on ${\mathcal M}^2$ is relevant because it does not involve the fields $\psi_I$ and $\psi_{IJ}$, it is much more interesting to see the changes in the canonical analysis in the chain ${\mathcal M}^2 \longrightarrow {\mathcal M}^3 \longrightarrow {\mathcal M}^4 \longrightarrow {\mathcal M}^5 \longrightarrow \cdots $. As will be clear in the lines below, the canonical analyses of the theory for $n\geq 5$ are very similar to the structure in $n=4$, with minor changes. However, the Hamiltonian descriptions in $n=2$ and in $n=3$ are very different from the cases $n\geq 4$, that is why theses cases are explicitly reported in the Appendices. Moreover, the two-dimensional case will be very useful to introduce local degrees of freedom (see Sec. \ref{local}). Thus, the canonical analysis for the action (\ref{action}) on ${\mathcal M}^4$ and higher-dimensional spacetimes is given in what follows. The Hamiltonian form of the action (\ref{action}) is obtained through Dirac's method \cite{dirac}
\begin{eqnarray}\label{ham}
S &=& \int \left [ \pi^a\,_I {\dot e}^I\,_a + \pi^a\,_{IJ} {\dot \omega}^{IJ}\,_a + \Pi^{ab}\,_I {\dot T}^I\,_{ab} + \Pi^{ab}\,_{IJ} {\dot R}^{IJ}\,_{ab} -{\mathcal H} \right] d^n x, \nonumber\\
{\mathcal H} &=& \lambda^I g_I + \lambda^{IJ} G_{IJ} + \Lambda^I\,_a d^a\,_I + \Lambda^{IJ}\,_a D^a\,_{IJ} \nonumber\\
&& + u_I\,^{ab} C^I\,_{ab} + u_{IJ}\,^{ab} \gamma^{IJ}\,_{ab} + v_I\,^{abc} C^I\,_{abc} + v_{IJ}\,^{abc} \gamma^{IJ}\,_{abc},
\end{eqnarray}
where ${\mathcal H}$ is the extended Hamiltonian \cite{henneaux}. From this expression it follows that the canonical pairs are: $(e^I\,_a, \pi^b\,_J)$, $(\omega^{IJ}\,_a, \pi^b\,_{KL})$, $(T^I\,_{ab},\Pi^{cd}\,_J)$, and $(R^{IJ}\,_{ab}, \Pi^{cd}\,_{KL})$, which coordinate the extended phase space. The expressions for the momenta in terms of the original Lagrangian variables are
\begin{eqnarray}
\pi^a\,_I &:=& \frac{1}{(n-2)!} \varepsilon^{0 a b_1 \ldots b_{n-2}}
\phi_{I\, b_1 \ldots b_{n-2}},
\nonumber \\
\pi^a\,_{IJ} &:=& \frac{1}{(n-2)!} \varepsilon^{0 a b_1 \ldots b_{n-2}}
\phi_{IJ\, b_1 \ldots b_{n-2}},
\nonumber \\
\Pi^{ab}\,_I &:=& \frac12 \frac{(-1)^{n-3}}{(n-3)!} \varepsilon^{0 a b c_1 \ldots c_{n-3}}
\psi_{I\, c_1 \ldots c_{n-3}},
\nonumber \\
\Pi^{ab}\,_{IJ} &:=& \frac12 \frac{(-1)^{n-3}}{(n-3)!} \varepsilon^{0 a b c_1 \ldots c_{n-3}} \psi_{IJ\, c_1 \ldots c_{n-3}},
\end{eqnarray}
while the ones for the Lagrange multipliers are: $\lambda^I= - e^I\,_0$, $\lambda^{IJ}=- \omega^{IJ}\,_0$, $\Lambda^I\,_a= T^I\,_{0a}$, $\Lambda^{IJ}\,_a= R^{IJ}\,_{0a}$, and
\begin{eqnarray}\label{multipliers}
u_I\,^{ab} &=& - \frac12 \frac{1}{(n-3)!} \varepsilon^{0 a b c_1 \ldots c_{n-3}}
\phi_{I\, 0 c_1 \ldots c_{n-3}}, \nonumber \\
u_{IJ}\,^{ab} &=& - \frac12 \frac{1}{(n-3)!} \varepsilon^{0 ab c_1 \ldots c_{n-3}}
\phi_{IJ\, 0 c_1 \ldots c_{n-3}},
\nonumber \\
v_I\,^{a b c} &=& \left \{
\begin{array}{l}
\frac12 \varepsilon^{0abc} \psi_{I\,0} \quad \mbox{if} \quad n=4, \\
\frac12 \frac{1}{(n-4)!} \varepsilon^{0 a b c d_1 \ldots d_{n-4}} \psi_{I\, 0 d_1 \ldots d_{n-4}} \quad \mbox{if}\quad  n \geq 5,
\end{array} \right.
  \nonumber \\
v_{IJ}\,^{a b c} &=& \left \{
\begin{array}{l}
 - \frac12 \varepsilon^{0abc} \psi_{IJ\, 0} \quad \mbox{if} \quad n=4, \\
- \frac12 \frac{1}{(n-4)!}  \varepsilon^{0 a b c d_1 \ldots d_{n-4}}
\psi_{IJ\, 0 d_1 \ldots d_{n-4}} \quad \mbox{if} \quad n \geq 5,
\end{array} \right.
\end{eqnarray}
which impose the constraints
\begin{eqnarray}\label{constraints}
g_I &:=& {\cal D}_a \pi^a\,_I - \Pi^{ab}\,_J R^J\,_{Iab} \approx 0,  \nonumber\\
G_{IJ} &:=& {\mathcal D}_a \pi^a\,_{IJ} + \frac12 \left ( \pi^a\,_I e_{Ja} - \pi^a\,_J e_{Ia} \right ) + \frac12 \left ( \Pi^{ab}\,_I T_{Jab} - \Pi^{ab}\,_J T_{Iab} \right ) + \Pi^{ab}\,_{IK} R_J\,^K\,_{ab} - \Pi^{ab}\,_{JK} R_I\,^K\,_{ab} \approx 0, \nonumber\\
d^a\,_I &:=& \pi^a\,_I + 2 {\mathcal D}_b \Pi^{ab}\,_I \approx 0, \nonumber\\
D^a\,_{IJ} &:=& \pi^a\,_{IJ} +2 {\mathcal D}_b \Pi^{ab}\,_{IJ} +
\Pi^{ab}\,_I e_{Jb} - \Pi^{ab}\,_J e_{Ib} \approx 0, \nonumber\\
C^I\,_{ab} &:=& {\mathcal D}_a e^I\,_b - {\mathcal D}_b e^I\,_a - T^I\,_{ab} \approx 0, \nonumber\\
\gamma^{IJ}\,_{ab} &:=& \partial_a \omega^{IJ}\,_b - \partial_b \omega^{IJ}\,_a
+ \omega^I\,_{Ka} \omega^{KJ}\,_b - \omega^I\,_{Kb} \omega^{KJ}\,_a -
R^{IJ}\,_{ab} \approx 0, \nonumber\\
C^I\,_{abc} &:=& R^I\,_{J[ab} e^J\,_{c]} - {\mathcal D}_{[a} T^I\,_{bc]} \approx 0, \nonumber\\
\gamma^{IJ}\,_{abc} &:=& {\mathcal D}_{[a} R^{IJ}\,_{bc]} \approx 0,
\end{eqnarray}
where ${\mathcal D}$ is the covariant derivative compute with respect to $\omega^I\,_{Ja}$.

A straightforward computation shows that the evolution of the constraints (\ref{constraints}) gives no additional ones. The constraints are smeared with test fields
\begin{eqnarray}
g(a) &=& \int {a}^I g_I, \quad  G(U)= \int {U}^{I J} G_{I J}, \quad
d(b) = \int {b}^{a I} d_{a I}, \quad D (V) = \int {V}^{a I J } D_{a I J}, \nonumber\\
f(c) &=& \int {c}^{I a b} C_{I a b}, \quad h (d) = \int {d}^{I J a b} \gamma_{I J a b}, \quad k(f) = \int {f}^{I a b c} C_{I a b c}, \quad l(g) = \int {g}^{I J a b c}
\gamma_{I J a b c},
\end{eqnarray}
to compute their Poisson algebra
\begin{eqnarray}\label{algebra}
\{ D(V), g(a) \} &=& d(V \cdot a), \qquad (V \cdot a)_{a K}
= {V}_{a K L} {a}^L;
\nonumber
\\
\{ G (U), g(a) \} &=& g(U \cdot a), \qquad (U \cdot a)_K =
{U}_{K L} {a}^L;
\nonumber
\\
\{ G(U), d(b) \} &=& d( U \cdot b ), \qquad (U \cdot b)_{a J}=
{U}_{J K} {b}_{a K};
\nonumber
\\
\{ G(U), D(V) \} &=& D( [U, V ] ), \qquad
[U, V ]_{a K J} = {U_{ K}}^L V_{ a L J}- {U_{ J}}^L V_{ a L K};
\nonumber
\\
\{ G(U_1), G(U_2) \} &=& G([U_1, U_2]), \qquad
[U_1, U_2]_{I K}= {U_{1 I}}^L U_{2 L K}-{U_{1 K}}^L U_{2 L I};
\nonumber
\\
\{ g(a), f(c) \} &=& h(c \cdot a), \qquad
(c \cdot a)_{I J a b}= \frac12(c_{ I a b} a_{ J}-c_{ J a b} a_{ I});
\nonumber
\\
\{ G(U), f(c) \} &=& f(U \cdot c), \qquad
 (U \cdot c)_{J a b} = {U_{ J}}^K c_{ K a b};
 \nonumber
 \\
 \{ G(U), h(d) \} &=& h([U, d]), \qquad
 [U, d]_{J L a b} = {U_{ J}}^K d_{ K L a b}-{U_{ L}}^K d_{ K J a b};
 \nonumber
 \\
\{ g(a), k(f) \} &=& l([a, f]), \qquad
[a, f]_{I J a b c} = \frac12 [a_{ I} f_{J a b c}-a_{ J} f_{ I a b c}];
\nonumber
\\
\{ G(U), k(f) \} &=& k(U \cdot f), \qquad
(U \cdot f)_{J a b c}= {U_{ J}}^K f_{ K a b c};
\nonumber
\\
\{ G(U), l(g) \} &=& l([U, g]), \qquad
[U, g]_{J L a b c} = {U_{ J}}^K g_{ K L a b c}
-{U_{ L}}^K g_{ K J a b c};
\nonumber
\\
\{ D(V), k(f) \} &=& f(V \cdot f), \qquad
 (V \cdot f)_{I c d} = {V_{  b I}}^J  f_{ J b c d};
 \nonumber
 \\
 \{ l(g), D(V) \} &=& h([V, g]), \qquad
 [V, g]_{I L c d}= {V_{ b I}}^K g_{K L b c d}-
 {V_{ b L}}^K g_{ K I b c d};
 \nonumber
 \\
 \{k(f), d(b) \} &=& h([b, f]), \qquad
  [b, f]_{I J c d}= \frac12 [{{b}^b}_I f_{ J b c d}-{{b}^b}_J f_{ I b c d}],
\end{eqnarray}
and the Poisson brackets that are not listed vanish strongly. Therefore, all the constraints in Eq. (\ref{constraints}) are first-class. If a naive counting of the number of local degrees of freedom were made without taking into account the reducibility of the constraints (\ref{constraints}), the outcome would be a negative number. In fact, as it is explained below, the reducibility pattern for the constraints (\ref{constraints}) is essentially the same as the one for the reducible constraints for pure BF theory in $n$-dimensional spacetimes. Therefore, before making the analysis of the reducibility of the first-class constraints for the current theory, it will be convenient to recall the corresponding analysis for pure BF theory living on $n$-dimensional spacetimes ${\mathcal M}^n$ for $n \geq 4$ \cite{cai95}. In that case, the reducible first-class constraints are
\begin{eqnarray}\label{bfred}
\varepsilon^{0 a_1 a_2 \dots a_{n-3} a_{n-2} a_{n-1}} F^i \,_{a_{n-2} a_{n-1}} (A) \approx 0, \quad i =1, \dots, {\rm dim}({\frak g}),
\end{eqnarray}
where $\left ( \frac12 F^i\,_{ab} (A) d x^a \wedge d x^b \right ) \otimes J_i$ is the curvature of the connection one-form $\left ( A^i\,_a d x^a \right ) \otimes J_i$, $J_i$ are the generators of the Lie algebra ${\frak g}$ and satisfy $[ J_i , J_j ] = c^k\,_{ij} J_k$.  Nevertheless, the constraints (\ref{bfred}) are not independent among themselves because the following chain of equations
\begin{eqnarray}
\varepsilon^{0 a_1 a_2 \ldots a_{n-3} a_{n-2} a_{n-1}} F^i\,_{a_{n-2} a_{n-1}} (A) & \approx & 0, \nonumber \\
\downarrow && \nonumber \\
\varepsilon^{0 a_1 a_2 \ldots a_{n-3} a_{n-2} a_{n-1}} \nabla_{a_{n-3}} F^i\,_{a_{n-2} a_{n-1}} (A) & = & 0, \nonumber \\
\downarrow && \nonumber \\
\vdots && \nonumber \\
\downarrow && \nonumber \\
\varepsilon^{0 a_1 a_2 \ldots a_{n-3} a_{n-2} a_{n-1}} \nabla_{a_1} \cdots \nabla_{a_{n-3}} F^i\,_{a_{n-2} a_{n-1}} (A) &=& 0,
\end{eqnarray}
can be obtained from them through the application of the internal covariant derivative $\nabla_a$ or, equivalently, renaming last expressions
\begin{eqnarray}\label{bf}
\phi^{i a_1 a_2 \ldots a_{n-3}} & \approx & 0, \nonumber\\
\downarrow && \nonumber \\
\nabla_{a_{n-3}} \phi^{i a_1 a_2 \ldots a_{n-3}} &=& 0, \nonumber \\
\downarrow && \nonumber \\
\vdots && \nonumber\\
\downarrow && \nonumber \\
\nabla_{a_1} \cdots \nabla_{a_{n-3}} \phi^{i a_1 a_2 \ldots a_{n-3}} &=& 0,
\end{eqnarray}
which are totally antisymmetric in the free indices. With the help of the diagram (\ref{bf}), the counting of the number of local degrees of freedom for pure BF theory is straightforward. The key point to do that is to realize that the equations in the first row in Eq. (\ref{bf}) correspond to the first-class constraints, the equations in the second row correspond to reducibility equations for the equations in the first row, however, these reducibility equations are also {\it not} independent among themselves, they are linked by the equations in the third row, and the equations in the third row are also not independent among themselves because they are linked by the ones in the fourth row, and so on until reaching the last row. Therefore, the number of {\it independent} first-class constraints in Eq. (\ref{bfred}) is equal to the number of equations in the first row in Eq. (\ref{bf}) {\it minus} the number of equations in the second row {\it plus} the number of equations in the third row {\it minus} the number of equations in the fourth row {\it plus} ... , etc., alternating the sign in the terms of the series until counting the number of equations in the last row with its corresponding sign. The result is that the number of independent first-class constraints in Eq. (\ref{bfred}) is $(n-2) \times {\rm dim}({\frak g})$ which must be added to ${\rm dim}({\frak g})$ equations contained in the Gauss law. Consequently, the total number of independent first-class constraints is $(n-1) \times {\rm dim}({\frak g})$ so that the number of local degrees of freedom in the configuration space is $\frac12 \left \{ 2 \left [ (n-1) \times {\rm dim}({\frak g}) \right ] - 2 \left [ (n-1) \times {\rm dim}({\frak g})\right ] \right \} =0 $, which means that the theory is topological. So much for pure BF theory.

Coming back to the theory studied in this paper, here it is also required to know the number of independent first-class constraints in Eq. (\ref{constraints}). This number can be obtained by first noting that the number of {\it independent} first-class constraints in $g_I$, $G_{IJ}$, $d^a\,_I$, and $D^a\,_{IJ}$ is simply equal to the number of these constraints minus their number of reducibility equations $N(g_I) + N(G_{IJ})$,
\begin{eqnarray}
N (g_I) + N(G_{IJ}) + N(d^a\,_I) + N(D^a\,_{IJ}) - N(g_I)- N(G_{IJ}) = N(d^a\,_I) + N(D^a\,_{IJ}),
\end{eqnarray}
where $N(g_I)$ denotes the number of equations in $g_I$ and so on; the
reducibility equations among $g_I$, $G_{IJ}$, $d^a\,_I$, and $D^a\,_{IJ}$ are obtained by applying the operator ${\mathcal D}_a$ to the constraints $d^a\,_I$ and $D^a\,_{IJ}$. What remains is to know the number of {\it independent} first-class constraints among the remaining ones
\begin{equation}\label{set2}
{C^I}_{a b} \approx 0, \quad {{\gamma^I}_J}_{a b} \approx 0,
\quad {{C^I}}_{a b c} \approx 0, \quad {{\gamma^I}_J}_{a b c} \approx 0,
\end{equation}
which are equivalent to
\begin{eqnarray}\label{sss}
&& \varepsilon^{0 a_1 \ldots a_{n-2} a_{n-1}} {C^I}_{a_{n-2} a_{n-1}} \approx 0, \quad
\varepsilon^{0 a_1 \ldots a_{n-2} a_{n-1}} {{\gamma^I}_J}_{a_{n-2} a_{n-1}} \approx 0, \nonumber
\\
& & \varepsilon^{0 a_1 \ldots a_{n-3} a_{n-2} a_{n-1}} {{C^I}}_{a_{n-3} a_{n-2} a_{n-1}} \approx 0, \quad
\varepsilon^{0 a_1 \ldots a_{n-3} a_{n-2} a_{n-1}} {{\gamma^I}_J}_{a_{n-3} a_{n-2} a_{n-1}} \approx 0,
\end{eqnarray}
or, renaming (\ref{sss})
\begin{equation}\label{xxx}
{\Phi^I}^{ a_1 \ldots a_{n-3} } \approx 0, \quad {{\Phi^I}_J}^{a_1 \ldots a_{n-3} } \approx 0, \quad
{\Psi^I}^{ a_1 \ldots a_{n-4} } \approx 0, \quad {{\Psi^I}_J}^{a_1 \ldots a_{n-4} } \approx 0,
\end{equation}
which are totally antisymmetric in the free indices. The constraints (\ref{xxx}) behave as constraints (\ref{bfred}) do for BF theory, i.e., repeatedly applying the operator ${\mathcal D}_a$ to Eqs. (\ref{xxx}), a chain of reducibility equations arises
\begin{center}
\begin{eqnarray}
\begin{array}{llll}
 {\Phi^I}^{ a_1 \ldots a_{n-3} } \approx 0, &  {{\Phi^I}_J}^{a_1 \ldots a_{n-3} } \approx 0, &
 {\Psi^I}^{ a_1 \ldots a_{n-4} } \approx 0, &  {{\Psi^I}_J}^{a_1 \ldots a_{n-4} } \approx 0,
 \nonumber
\\
 {\Phi^I}^{ a_1 \ldots a_{n-4} }=0, &  {{\Phi^I}_J}^{a_1 \ldots a_{n-4} }=0, &
 {\Psi^I}^{ a_1 \ldots a_{n-5} }=0, &  {{\Psi^I}_J}^{a_1 \ldots a_{n-5} }=0,
\nonumber
 \\
 \vdots & \vdots & \vdots & \vdots
\\
\nonumber
 {\Phi^I}^{ a_1 \ldots a_{n-3-k} }=0, &  {{\Phi^I}_J}^{a_1 \ldots a_{n-3-k} }=0, &
 {\Psi^I}^{ a_1 \ldots a_{n-4-k} }=0, &  {{\Psi^I}_J}^{a_1 \ldots a_{n-4-k} }=0,
 \nonumber
 \\
  \vdots & \vdots & \vdots & \vdots
 \nonumber
 \\
 {\Phi^I}^{ a_1  a_2 }=0, &  {{\Phi^I}_J}^{a_1 a_2 }=0, &
 {\Psi^I}^{ a_1 }=0, &  {{\Psi^I}_J}^{a_1}=0,
 \nonumber
 \\
 {\Phi^I}^{ a_1 }=0, &  {{\Phi^I}_J}^{a_1}=0, &
 \Psi^I=0, &  {\Psi^I}_J=0,
 \nonumber
 \\
 {\Phi^I}=0, &  {\Phi^I}_J=0. & &
\end{array}
\end{eqnarray}
\end{center}
By using this chain, it is possible to make the counting of {\it independent} first-class constraints in Eq. (\ref{set2}) by adding (with the corresponding sign) the numbers contained in each row of the following diagram
\begin{center}
\begin{eqnarray}
\begin{array}{lllll}
(+)^{0} & N({\Phi^I}^{ a_1 \ldots a_{n-3} }), &  N({{\Phi^I}_J}^{a_1 \ldots a_{n-3} }), & N({\Psi^I}^{ a_1 \ldots a_{n-4} }), &  N({{\Psi^I}_J}^{a_1 \ldots a_{n-4} }), \nonumber \\
(-)^{1} & N({\Phi^I}^{ a_1 \ldots a_{n-4} }), &  N({{\Phi^I}_J}^{a_1 \ldots i_{n-4} }), & N({\Psi^I}^{ a_1 \ldots a_{n-5} }), &  N({{\Psi^I}_J}^{a_1 \ldots a_{n-5} }), \nonumber\\
\vdots & \vdots & \vdots & \vdots & \vdots \\ \nonumber
(-)^k & N({\Phi^I}^{ a_1 \ldots a_{n-3-k} }), &  N({{\Phi^I}_J}^{a_1 \ldots a_{n-3-k} }), & N({\Psi^I}^{ a_1 \ldots a_{n-4-k} }), &  N({{\Psi^I}_J}^{a_1 \ldots a_{n-4-k} }), \nonumber \\ \vdots & \vdots & \vdots & \vdots & \vdots \nonumber \\
(-)^{n-5} & N({\Phi^I}^{ a_1  a_2 }), &  N({{\Phi^I}_J}^{a_1 a_2 }), & N({\Psi^I}^{ a_1 }), &  N({{\Psi^I}_J}^{a_1}), \nonumber\\
(-)^{n-4} & N({\Phi^I}^{ a_1 }), &  N({{\Phi^I}_J}^{a_1}), & N(\Psi^I), &  N({\Psi^I}_J), \nonumber\\
(-)^{n-3} & N({\Phi^I}), &  N({\Phi^I}_J). & &
\end{array}
\end{eqnarray}
\end{center}
Taking into account the signs, it is clear that the third and fourth numbers of the first row are canceled by the first and second numbers of the second row, the third and fourth numbers of the second row are canceled by the first and second terms of the third row, etc.; in such a way that the two numbers of the last row are canceled by the third and fourth terms of the penultimate row. Therefore, the number of independent first-class constraints in the set (\ref{set2}) is simply $N(\Phi^{I a_1 \ldots a_{n-3}}) + N (\Phi^I\,_J\,^{a_1 \ldots a_{n-3}})= N(C^I\,_{ab})+ N(\gamma^I\,_{J\,ab})$, i.e., the information needed to get the right number of independent constraints is encoded in the number of the constraints $ {C^I}_{a b}$ and ${{\gamma^I}_J}_{a b}$ only. Using this result and the previous one it is concluded that the total number of independent first-class constraints in Eq. (\ref{constraints}) is $N (d^a\,_I) + N(D^a\,_{IJ}) + N(C^I\,_{ab})+ N(\gamma^{IJ}\,_{ab})$. Therefore, the number of local degrees of freedom in the configuration space is
\begin{eqnarray}
\frac12 \left \{ 2 \left [ N(e^I\,_a) + N(\omega^{IJ}\,_a) + N(T^I\,_{ab}) + N(R^{IJ}\,_{ab}) \right ] - 2 \left [ N (d^a\,_I) + N(D^a\,_{IJ}) + N(C^I\,_{ab})+ N(\gamma^{IJ}\,_{ab}) \right ] \right \} =0,
\end{eqnarray}
so the theory defined by the action (\ref{action}) is topological. Note that the number of variables in $e^I\,_a$, denoted by $N(e^I\,_a)$, is equal to the number of equations in $d^I\,_a$, denoted by $N(d^I\,_a)$, and so on.

Finally, for spacetimes ${\mathcal M}^n$ with $n \geq 3$, it is possible to use the equations of motion contained in the fifth and sixth rows of Eq. (\ref{emotion}) and insert back into the action (\ref{action}) the expressions for $\phi_I$ and $\phi_{IJ}$ in terms of the fields $\omega^I\,_J$, $\psi_I$, $\psi_{IJ}$, and $e^I$ to seek for an equivalent form for the action principle. However, this leads to the following result
\begin{eqnarray}\label{fea}
\int_{\mathcal{M}^n} (-1)^{n-3} d \left [ \psi_I \wedge \left ( T^I - D e^I \right ) + \psi_{IJ} \wedge \left ( R^{IJ} - R^{IJ} (\omega) \right ) \right ].
\end{eqnarray}
The fact that the Lagrangian $n$-form of the action (\ref{action}) can be written as the differential of a $(n-1)$-form is, in a certain sense, analog to the fact that products of curvature of type $F(A)$ wedge $F(A)$ are equal to the differential of the Chern-Simons Lagrangians. The difference between them and the current theory lies in the fact that in those cases the result is obtained without using any equations of motion while here some equations of motion were used to get (\ref{fea}).

\section{Particular topological field theories and generalizations}\label{general}

\subsection{Particular theories}\label{particular}
By plugging $T^I=0$ and $R^{IJ}=0$ into the Lagrangian action (\ref{action}) leads to the field theory
\begin{eqnarray}\label{pt}
S[\omega^I\,_J, e^I, \phi_I, \phi_{IJ}] &=& \int_{\mathcal{M}^n} \left [ \phi_I  \wedge \left ( d e^I + \omega^I\,_J \wedge e^J  \right ) + \phi_{IJ} \wedge \left ( d \omega^{IJ} + \omega^I\,_K \wedge \omega^{KJ} \right ) \right],
\end{eqnarray}
which is well defined for spacetime manifolds ${\mathcal M}^n$ with $n\geq 2$. Note that this action principle contains also BF theory with $SO(n)$ or $SO(n-1,1)$ structure groups as particular cases. Using the results of Sec. \ref{theory}, it is easy to show that this action defines a topological field theory too \footnote{ As far as we know, this is the first time the theory (\ref{pt}) is reported. With respect to this, in Ref. \cite{pibe} an action principle for a two-dimensional theory was reported in  Eq. (A1), which contains the action (19) as part of the action reported there. Nevertheless, there the idea was to report the action (A1) as a whole thing
defining a two-dimensional topological field theory and an analysis of their
parts was not carried out. More precisely, the action (19) was not studied by
itself (alone) in Ref. \cite{pibe}. Furthermore, it was not realized there that the action (19) was topological as it is done in this paper. Also, action (19) is defined for spacetimes with dimension 2 and higher, not just two-dimensional ones.}. The Hamiltonian form for the action (\ref{pt}), obtained through Dirac's method, is
\begin{eqnarray}
S &=& \int \left [ \pi^a\,_I {\dot e}^I\,_a + \pi^a\,_{IJ} {\dot \omega}^{IJ}\,_a
- {\mathcal H} \right] d^n x, \nonumber\\
{\mathcal H} &=& \lambda^I g_I + \lambda^{IJ} G_{IJ} + u_I\,^{ab} C^I\,_{ab} + u^{IJ}\,_{ab} \gamma^{IJ}\,_{ab},
\end{eqnarray}
where now
\begin{eqnarray}\label{hampt}
g_I &:=& {\mathcal D}_a \pi^a\,_I  \approx 0,  \nonumber\\
G_{IJ} &:=& {\mathcal D}_a \pi^a\,_{IJ} + \frac12 \left ( \pi^a\,_I e_{Ja} - \pi^a\,_J e_{Ia} \right ) \approx 0, \nonumber\\
C^I\,_{ab} &:=& {\mathcal D}_a e^I\,_b - {\mathcal D}_b e^I\,_a  \approx 0, \nonumber\\
\gamma^{IJ}\,_{ab} &:=& \partial_a \omega^{IJ}\,_b - \partial_b \omega^{IJ}\,_a
+ \omega^I\,_{Ka} \omega^{KJ}\,_b - \omega^I\,_{Kb} \omega^{KJ}\,_a  \approx 0.
\end{eqnarray}
A straightforward computation shows that the algebra of constraints closes so that the constraints (\ref{hampt}) are first class. Alternatively, from the constraint algebra of the Hamiltonian analyses for the two-dimensional, three-dimensional, and the generic theory developed in Appendices \ref{aa} and \ref{ab}, and in Sec. \ref{theory}, respectively, it follows that the constraints $d^a\,_I$, $D^a\,_{IJ}$, $C^I\,_{abc}$, and $\gamma^{IJ}\,_{abc}$ can be dropped from the Hamiltonian form (\ref{ham}) in such a way that the constraints of the smaller set also close, so they are first-class constraints too. The constraints of Eq. (\ref{hampt}) are irreducible for two-dimensional and three-dimensional spacetime manifolds while the constraints $C^I\,_{ab}$ and $\gamma^{IJ}\,_{ab}$ become reducible for $n \geq 4$. Following the same procedure made for the analysis of the reducibility of the constraints performed in Sec. \ref{theory} it follows that the number of independent first-class constraints among $C^I\,_{ab}$ and $\gamma^{IJ}\,_{ab}$ is $(n-2)\times N(g_I) + (n-2) \times N(G_{IJ})$. On the other hand, the constraints $g_I$ and $G_{IJ}$ are always irreducible. Thus, the total number of independent first-class constraints in (\ref{hampt}) is
\begin{eqnarray}
(n-2)\times N(g_I) + (n-2) \times N(G_{IJ}) + N(g_I) + N(G_{IJ})= (n-1)\times N(g_I) + (n-1) \times N(G_{IJ}),
\end{eqnarray}
which implies that the number of local degrees of freedom in the configuration space is
\begin{eqnarray}
\frac12 \left \{ 2 \left [ N(e^I\,_a) + N(\omega^{IJ}\,_a)\right ] - 2 \left [
(n-1)\times N(g_I) + (n-1) \times N(G_{IJ})
\right ] \right \} =0,
\end{eqnarray}
which means that the theory (\ref{pt}) is topological.

\subsection{Generalizations}\label{beauty}
In the action principles (\ref{action}), (\ref{accion}), and (\ref{pt}), studied in Secs.  \ref{theory} and \ref{particular}, the group of local orthogonal rotations was taken to be $SO(n)$ or $SO(n-1,1)$, i.e., the one that corresponds naturally to $n$-dimensional spacetime manifolds. Nevertheless, it follows immediately from their canonical analyses that there is no need of restricting the analysis to that group in spite of the fact that the theories are defined on $n$-dimensional spacetimes. In fact, the group can be $SO(m)$ or $SO(m-1,1)$ with $m \neq n$. If this were allowed, the counting of the degrees of freedom would be left unaltered and the various theories would remain topological, the challenge would be the interpretation of the various fields involved only (see Sec. \ref{local}). This feature is indeed very interesting and has implications on at least the following two issues:
\begin{enumerate}
\item
First of all, it is well-known that the freedom in the choice of the dimension of the group $SO(m)$ or $SO(m-1,1)$ can be used to introduce ``matter fields'' that will interact with those degrees of freedom naturally living on $n$-dimensional spacetime manifolds (see Sec. \ref{local} for a concrete implementation of this idea, for instance).
\item
On the other hand, if $m > n$ then the theories studied along this paper can be naturally coupled to other theories living on higher-dimensional spacetime manifolds ${\mathcal M}^m$ having $SO(m)$ or $SO(m-1,1)$ as the group of local orthogonal rotations. From this point of view, the former theories might be interpreted as extended objects (strings, membranes, etc.) that can be naturally coupled and acting as sources for the fields of geometric theories living on higher-dimensional spacetimes ${\mathcal M}^m$. In fact, following this way of thinking, in the context of nonperturbative quantum gravity, various two-dimensional topological field theories allowing the existence of ``tetrad fields'' (and therefore allowing either $SO(3,1)$ or $SO(4)$ groups) living on a two-dimensional spacetime have been constructed out and coupled to four-dimensional BF theories \cite{pibe}. The last models are conceptually different from and technically equal to those analyzed in Ref. \cite{baezp} where the coupling of $(n-3)$-dimensional membranes to $n$-dimensional BF theory defined for a large class of structure groups was studied. On this matter see also Refs. \cite{winspibe,winston} as well as the previous results contained in Refs. \cite{karim,carlip}.
\end{enumerate}

\section{Adding local degrees of freedom}\label{local}
Up to now, the paper has been focused in the analysis of various diffeomorphism invariant topological field theories that are {\it per se} interesting enough. Nevertheless, one might wonder about the relationship between them and theories with local degrees of freedom, such as general relativity or modifications of it, for instance. To be precise, the question is how to build theories with local degrees of freedom from the topological theories discussed in Secs. \ref{theory} and \ref{general}. This is the issue studied in this section.

The easiest way of modifying the action principles discussed up to now, in order to build field theories with local degrees of freedom, is to use the canonical analyses performed in Secs. \ref{theory} and \ref{general}, and in the appendices. From them it follows that there are, essentially, three parameters at hand to generate local degrees of freedom: the number of phase space variables, the number of constraints, and the number of reducibility equations. The main idea developed here is that by means of an appropriate handling of these parameters it is possible to get theories with a nonvanishing number of physical degrees of freedom. This process implies to get rid of the topological nature of the original theory to allow local excitations to emerge. Even though this way of generating local degrees of freedom is right from the Hamiltonian viewpoint, the challenge is that this procedure is covariant in the sense that it is associated to a Lagrangian action principle.

In what follows it is shown that this is indeed possible and it is illustrated in two-dimensional spacetimes using the results contained in Appendix \ref{aa}. More precisely, this idea is implemented through the following steps:
\begin{enumerate}
\item
Take the Hamiltonian action principle given in Eq. (\ref{hamthe2}).
\item
Take $SO(2,1)$ or $SO(3)$ as the internal gauge group, i.e., even though the spacetime is two-dimensional, the gauge group is not $SO(1,1)$ or $SO(2)$ but $SO(2,1)$ or $SO(3)$. It has been shown in Sect. \ref{general} that this is indeed allowed without destroying the topological character of the theory.
\item
Add the equation among the momenta
\begin{eqnarray}
\pi_{IJ} - a\, \varepsilon_{IJK} \pi^K =0,
\end{eqnarray}
where $a$ is a constant.
\end{enumerate}
The key point is to realize that this relationship corresponds to a reducibility equation
\begin{eqnarray}
D_{IJ} - a\, \varepsilon_{IJK} d^K=0,
\end{eqnarray}
among the constraints $d^I$ and $D_{IJ}$. In this way, the new theory has more reducibility equations than the original one (\ref{hamthe2}), the counting is not balanced and the new theory is {\it not} topological anymore, it has
\begin{eqnarray}
1/2 [ 2(3+3) - 2(3+3+3+3- 3-3-3) ] = 1/2[ 2(6)- 2(3)]=3,
\end{eqnarray}
local degrees of freedom. Even though this Hamiltonian way of generating degrees of freedom is correct, it remains to see that it corresponds to a Lagrangian action. This is really so, the corresponding Lagrangian action is
\begin{eqnarray}\label{agl}
S[\omega^I\,_J, e^I, T^I, R^I\,_J, \phi_I] &=& \int_{\mathcal{M}^2} \left [ \phi_I  \left ( d e^I + \omega^I\,_J \wedge e^J - T^I \right ) + a\, \varepsilon_{IJK} \phi^K \left ( d \omega^{IJ} + \omega^I\,_M \wedge \omega^{MJ} - R^{IJ} \right ) \right].
\end{eqnarray}
The field theory of the action principle (\ref{agl}) is two-dimensional gravity coupled to additional fields. The contact with two-dimensional gravity is made through the usual identification $I,J={\hat a},2$ with ${\hat a}=0,1$ (see Refs. \cite{cham,cham2})
\begin{eqnarray}
\omega^{{\hat a}{\hat b}} &=& \Omega^{{\hat a}{\hat b}}, \quad \omega^{{\hat a}2}= \vartheta^{\hat a}, \nonumber\\
(\phi^I) &=& \left ( \phi^{\hat a}, \phi  \right ),
\end{eqnarray}
where $\Omega^{{\hat a}{\hat b}}$ is a two-dimensional Lorentz connection and $\vartheta^{\hat a}$ is a two-dimensional local Lorentz frame; two-dimensional Lorentz indices ${\widehat a}, {\widehat b}$ are raised and lowered with the metric $\left (\eta_{{\widehat a}, {\widehat b}} \right )= \mbox{diag}\,\, (\sigma, +1)$. Therefore, action (\ref{agl}) becomes
\begin{eqnarray}\label{lasta}
S &=& \int_{\mathcal{M}^2} \left [ \phi_{\hat a} \, d_{\Omega} e^{\hat a} + \phi_{\hat a} \, \vartheta^{\hat a} \wedge e^2 - \phi_{\hat a} T^{\hat a} + \phi \left ( d e^2 - \vartheta_{\hat a} \wedge e^{\hat a} - T^2 \right ) \right. \nonumber\\
&& \left. + a\, \phi \, \varepsilon_{{\hat a}{\hat b}} \, R^{{\hat a}{\hat b}}({\Omega}) - a\, \phi \, \varepsilon_{{\hat a}{\hat b}} \, \vartheta^{\hat a} \wedge \vartheta^{\hat b} - a\, \phi \, \varepsilon_{{\hat a}{\hat b}} R^{{\hat a}{\hat b}} - 2 a\, \varepsilon_{{\hat a} {\hat b}}\, \phi^{\hat b}\, d_{\Omega} \vartheta^{\hat a} + 2 a\,  \varepsilon_{{\hat a}{\hat b}}\, \phi^{\hat b}\, R^{{\hat a} 2} \right],
\end{eqnarray}
with $\varepsilon_{{\hat a}{\hat b}2}= \varepsilon_{{\hat a}{\hat b}}$ and $R^{{\hat a}{\hat b}} (\Omega) = d \Omega^{{\hat a}{\hat b}} + \Omega^{\hat a}\,_{\hat c} \wedge \Omega^{{\hat c}{\hat b}}$. From this expression it is clearly observed that the first, second, and fourth terms in the second row correspond to two-dimensional gravity (see Eq. (2.12) of Ref. \onlinecite{cham2}). The additional terms in (\ref{lasta}) give the explicit nonminimal couplings of the matter fields to gravity. Even though the terms in the action (\ref{lasta}) show the nature of the dynamical fields through their couplings, the particular dynamics of one of the matter fields involved can be illustrated even more from the equations of motion that follow from the variation of the action (\ref{agl}) with respect to $e^I$
\begin{eqnarray}\label{mmmh}
&& d \phi^{\hat a} + \vartheta^{\hat a}\, \phi + \Omega^{\hat a}\,_{\hat b} \phi^{\hat b} = 0, \nonumber\\
&& \phi_{\hat a} = \vartheta_{\hat a}\,^{\alpha} \partial_{\alpha} \phi.
\end{eqnarray}
In fact, plugging the equation of the second row into the one of the first row of (\ref{mmmh}) leads to
\begin{eqnarray}\label{klein}
 g^{\alpha\beta} \partial_{\alpha} \partial_{\beta} \phi + \vartheta_{\hat a}\,^{\alpha} \partial_{\alpha} \vartheta^{\beta {\hat a}} \partial_{\beta} \phi +  2 \phi + \vartheta_{\hat a}\,^{\alpha} \Omega^{\hat a}\,_{\hat b}\,_{\alpha} \vartheta^{\beta{\hat b}} \partial_{\beta} \phi =0,
\end{eqnarray}
where $g^{\alpha\beta}= \vartheta_{\hat a}\,^{\alpha} \vartheta^{{\hat a} \beta}$ is the inverse of the induced two-dimensional metric $g_{\alpha\beta}= \vartheta^{\hat a }\,_{\alpha} \vartheta^{\hat b}\,_{\beta} \eta_{{\hat a}{\hat b}}$. The expression for $\Omega^{{\hat a}{\hat b}}\,_{\mu}$ can be obtained from the variation of the action (\ref{agl}) with respect to $\phi_I$, it is of the form $\Omega^{{\hat a}{\hat b}}\,_{\mu} = \Gamma^{{\hat a}{\hat b}}\,_{\mu} + S^{{\hat a}{\hat b}}\,_{\mu}$ where $\Gamma^{{\hat a}{\hat b}}\,_{\mu}$ is the spin connection and $S^{{\hat a}{\hat b}}\,_{\mu}$ includes the contribution of matter fields. Note that Eq. (\ref{klein}) is an extension of Eq. (2.28) of Ref. \onlinecite{cham3} because in that case $\Omega^{{\hat a}{\hat b}}\,_{\mu}$ does not include $S^{{\hat a}{\hat b}}\,_{\mu}$.

Furthermore, by defining the fields
\begin{eqnarray}
^{\gamma}A^{IJ} &:=& \omega^{IJ} + \gamma \varepsilon^{IJ}\,_K e^K, \nonumber\\
^{\beta}A^{IJ} &:=& \omega^{IJ} + \beta \varepsilon^{IJ}\,_K e^K,
\end{eqnarray}
with $\gamma -\beta \neq 0$, the meaning of the action (\ref{agl}) becomes clearer due to the fact it can be cast in the equivalent form
\begin{eqnarray}\label{accion2}
S[^{\gamma}A^{IJ}, ^{\beta}A^{IJ} , T^I, R^I\,_J, \phi_I] &=& \int_{\mathcal{M}^2} \left [ \frac{\beta}{2 \gamma} \frac{1}{\beta - \gamma} \varepsilon_{IJK} \phi^{K} F^{IJ} (^{\gamma}A) - \frac{\gamma}{2 \beta} \frac{1}{\beta - \gamma} \varepsilon_{IJK} \phi^{K} F^{IJ} (^{\beta}A) \right. \nonumber\\
&& \left.  - \frac{\gamma + \beta}{2 \gamma \beta} \varepsilon_{IJK} \phi^K R^{IJ} - \phi_I T^I \right],
\end{eqnarray}
that involves two interacting BF theories sharing the ``$B$ field''; the constant $a$ in (\ref{agl}) is related to the constants $\gamma$ and $\beta$ through $a= \frac{\gamma+\beta}{2\gamma\beta}$, which has been chosen in order to eliminate the quadratic terms in both connections $^{\gamma}A$ and $^{\beta}A$ that appear when the quadratics terms are recollected. Of course, it would be nice to try to get an equivalent form for the action
(\ref{agl}), (\ref{lasta}) or (\ref{accion2}) of the model containing the true (physical) degrees of freedom only. However, that is not the point here; the point is to show that idea putting forward in this paper works, namely, that it is possible to build theories with local degrees of freedom from the original topological theory by means of a suitable modification that destroys its topological nature and that allows the emerging of local excitations.

It is worth noting that the local excitations have arisen essentially by establishing some additional relations among the variables involved that were not present in the original topological field theory. Note, however, that the constraints were directly imposed on the fields involved in the model ($\phi_I$ and $\phi_{IJ}$, in this case). Of course it is also possible to incorporate the constraints on the fields by introducing more auxiliary fields that impose these constraints, which would be much more in the spirit of the relationship between BF gravity and pure BF theory \cite{jmp1977,penrose,capo,cqg1999,cqg1999b,rob,cqgl2001}.

\section{Concluding Remarks}\label{fin}
A generalization of part of the results contained in Ref. \cite{cuesta} concerning Cartan's equations and Bianchi identities was presented. Contrary to the previous work, the current action principles for diffeomorphism invariant topological field theories hold for arbitrary $n$-dimensional spacetime manifolds and involve auxiliary fields. It is worthwhile to mention that the connection $\omega^I\,_J $ involved in the action principles (\ref{action}) and (\ref{accion}) is not flat; its curvature is equal to the two-form field $R^{IJ}$. This is a major difference between theories (\ref{action}) and (\ref{accion}) of this paper and pure BF theories. It was also shown that it is possible to use this theoretical framework to build a two-dimensional field theory with local degrees of freedom by imposing additional restrictions on the fields involved, destroying the topological nature of the original theory. It would be interesting to apply the same strategy in four-dimensional spacetimes to find alternative formulations for general relativity or modifications (generalizations) of it, just starting from the action principle (\ref{action}) and constraining the fields suitably. This is left for future work. Because of the fact that the framework developed in Secs. \ref{theory} and \ref{general} is quite generic, it is natural to expect that it can also be applied to analyze general relativity in arbitrary finite-dimensional spacetime manifolds, in the sense of \cite{fkp}.

There are various topics that were not touched in the paper, but they deserve also to be explored, among these: (1) the interpretation of the auxiliary fields involved, (2) the relationship between this approach and the one of Ref. \cite{cuesta} if the analysis of this paper is restricted to four-dimensional spacetimes, (3) the possible relationship between the various topological field theories reported in the paper with other topological theories, and (4) the inclusion of fermion fields in the current theoretical framework.

\section*{ACKNOWLEDGMENTS}
This work was supported in part by CONACYT, M\'exico, Grant No. 56159-F and No. 47211-F. J.D. Vergara acknowledges the support from DGAPA-UNAM Grant No. IN109107. V. Cuesta and M. Vel\'azquez acknowledge the financial support from CONACYT.

\appendix

\section{Two-dimensional theory}\label{aa}
The equations of motion that follow from the variation of the action (\ref{accion}) with respect to the independent fields are
\begin{eqnarray}\label{emotion1}
&& \delta \phi_I:  d e^I + \omega^I\,_J \wedge e^J - T^I =0, \nonumber\\
&& \delta \phi_{IJ}: d \omega^{IJ} + \omega^I\,_K \wedge \omega^{KJ} - R^{IJ} =0, \nonumber\\
&& \delta T^I: \phi_I =0, \nonumber\\
&& \delta R^{IJ}: \phi_{IJ}=0, \nonumber\\
&& \delta \omega^{IJ}: D\phi_{IJ} + \frac12 \left ( \phi_I e_J - \phi_J e_I \right ) = 0, \nonumber\\
&& \delta e^I: D \phi_I =0.
\end{eqnarray}
The Hamiltonian form of the action (\ref{accion}) is
\begin{eqnarray}\label{hamthe2}
S[e^I\,_1, \omega^{IJ}\,_1, \pi_I, \pi_{IJ}, \lambda^I, \lambda^{IJ}, \Lambda^I, \Lambda^{IJ} ] &=& \int \left [ \pi_I {\dot e}^I\,_1 + \pi_{IJ} {\dot \omega}^{IJ}\,_1 \right. \nonumber\\
&& \left.  - \lambda^I g_I - \lambda^{IJ} G_{IJ} - \Lambda^I d_I - \Lambda^{IJ} D_{IJ} \right ]d x^0 \wedge d x^1,
\end{eqnarray}
where
\begin{eqnarray}\label{const1}
g_I &:=& \partial_{x^1} \pi_I - \omega^K\,_{I1} \pi_K \approx 0, \nonumber\\
G_{IJ} &:=& \partial_{x^1} \pi_{IJ} - \omega^K\,_{I1} \pi_{KJ} -  \omega^K\,_{J1} \pi_{IK} + \frac12 \left ( \pi_I e_{J1} - \pi_J e_{I1} \right ) \approx 0, \nonumber\\
d_I &:=& \pi_I \approx 0, \nonumber\\
D_{IJ} &:=& \pi_{IJ} \approx 0,
\end{eqnarray}
and the following definition of variables has been made: $\pi_I:= \phi_I$, $\pi_{IJ}:= \phi_{IJ}$, $\lambda^I:= - e^I\,_0$, $\lambda^{IJ}:=-\omega^{IJ}\,_0$, $\Lambda^I:= T^I\,_{01}$, and $\Lambda^{IJ}:= R^{IJ}\,_{01}$. Smearing the constraints (\ref{const1}) with test fields whose indices have the corresponding symmetries of the constraints
\begin{eqnarray}
g(a):= \int d x^1 a^I g_I, \quad G(u):= \int d x^1 u^{IJ} G_{IJ}, \quad d(\alpha):= \int d x^1  \alpha^I \pi_I, \quad D(U):= \int d x^1 U^{IJ} \pi_{IJ},
\end{eqnarray}
and computing their Poisson brackets gives the nonvanishing ones
\begin{eqnarray}
\{ G(u), g(a) \} &=& g(u \cdot a),\nonumber\\
\{ D(U), g(a) \} &=& d (U \cdot a) ,\nonumber\\
\{ G(u), G(v) \} &=& G([u,v]),\nonumber\\
\{ G(u), d(\alpha) \} &=& d (u \cdot \alpha), \nonumber\\
\{ G(u), D(U) \} &=& D([u,U]),
\end{eqnarray}
where $(u \cdot a)^I := u^{IJ} a_J$, $(U \cdot a)^I := U^{IJ} a_J$, $[u,v]^{IJ}:= u^I\,_K v^{KJ}- u^J\,_K v^{KI}$, $(u \cdot \alpha)^I := u^{IJ} \alpha_J$, and $[u,U]^{IJ}:= u^I\,_K U^{KJ} - u^J\,_K U^{KI}$. So, the $2+1+2+1=6$ constraints in Eq. (\ref{const1}) are first class. However, they are reducible because of the $2+1=3$ reducibility equations
\begin{eqnarray}
g_I - \partial_{x^1} d_I + \omega^K\,_{I1} d_K &=& 0, \nonumber\\
G_{IJ}- \partial_{x^1} D_{IJ} + \omega^K\,_{I1} D_{KJ} +  \omega^K\,_{J1} D_{IK} - \frac12 \left ( d_I e_{J1} - d_J e_{I1} \right ) &=& 0.
\end{eqnarray}
Therefore, there are just $6-3=3$ independent first-class constraints in Eq. (\ref{const1}). because of the fact that there are $2+1=3$ configuration variables, the number of local degrees of freedom is $\frac12 \left [ 2 (3) - 2(6-3) \right ]=0$ and so the action (\ref{accion}) defines a topological field theory.

In this appendix the structure group was taken to be $SO(2)$ or $SO(1,1)$. Nevertheless, as it was explained in Sec. \ref{beauty}, the computation can be performed generically  and it turns out that the theory remains topological for the group $SO(m)$ or $SO(m-1,1)$ in spite of the fact that the theory lives in a two-dimensional spacetime. This comment about the structure group also applies to the analysis carried out in Appendix \ref{ab} where the structure group was taken to be $SO(3)$ or $SO(2,1)$.

\section{Three-dimensional theory}\label{ab}
The theory is defined by the action (\ref{action}) where the fields $\phi_I$ are three one-forms and $\phi_{IJ}$ are three one-forms while $\psi_I$ are three $0$-forms, and $\psi_{IJ}$ are three $0$-forms, respectively. The Hamiltonian form of the action (\ref{action}) is
\begin{eqnarray}
&& \int \left [ \pi^a\,_I {\dot e}^I\,_a + \pi^a\,_{IJ} {\dot \omega}^{IJ}\,_a + \Pi^{ab}\,_I {\dot T}^I\,_{ab} + \Pi^{ab}\,_{IJ} {\dot R}^{IJ}\,_{ab} \right. \nonumber\\
&& \left. - \lambda^I g_I - \lambda^{IJ} G_{IJ} - \Lambda^I\,_a d^a\,_I - \Lambda^{IJ}\,_a D^a\,_{IJ} - u_I h^I - u_{IJ} H^{IJ} \right ] d^3 x,
\end{eqnarray}
where the definitions (in terms of the original variables) of the momenta are: $\pi^a\,_I:= \varepsilon^{ab} \phi_{Ib}$, $\pi^a\,_{IJ}:= \varepsilon^{ab} \phi_{IJb}$, $\Pi^{ab}\,_I:= \frac12 \varepsilon^{ab} \psi_I$, $\Pi^{ab}\,_{IJ}:= \frac12 \varepsilon^{ab} \psi_{IJ}$ while the Lagrange multipliers $\lambda^I:= - e^I\,_0$, $\lambda^{IJ}:=-\omega^{IJ}\,_0$, $\Lambda^I\,_a:= T^I\,_{0a}$, $\Lambda^{IJ}\,_a:= R^{IJ}\,_{0a}$, $u_I := -\frac12 \phi_{I0}$, and $u_{IJ}:= - \frac12 \phi_{IJ0}$ impose the constraints
\begin{eqnarray}\label{const2}
g_I &:=& {\mathcal D}_a \pi^a\,_I - \Pi^{ab}\,_J R^J\,_{Iab} \approx 0,  \nonumber\\
G_{IJ} &:=& {\mathcal D}_a \pi^a\,_{IJ} + \frac12 \left ( \pi^a\,_I e_{Ja} - \pi^a\,_J e_{Ia} \right ) + \frac12 \left ( \Pi^{ab}\,_I T_{Jab} - \Pi^{ab}\,_J T_{Iab} \right ) + \Pi^{ab}\,_{IK} R_J\,^K\,_{ab} - \Pi^{ab}\,_{JK} R_I\,^K\,_{ab} \approx 0, \nonumber\\
d^a\,_I &:=& \pi^a\,_I + 2 {\mathcal D}_b \Pi^{ab}\,_I \approx 0, \nonumber\\
D^a\,_{IJ} &:=& \pi^a\,_{IJ} +2 {\mathcal D}_b \Pi^{ab}\,_{IJ} +
\Pi^{ab}\,_I e_{Jb} - \Pi^{ab}\,_J e_{Ib} \approx 0, \nonumber\\
h^I &:=& \varepsilon^{ab} \left ( {\mathcal D}_a e^I\,_b - {\mathcal D}_b e^I\,_a  - T^I\,_{ab} \right ) \approx 0, \nonumber\\
H^{IJ} &:=& \varepsilon^{ab} \left ( \partial_a \omega^{IJ}\,_b -  \partial_b \omega^{IJ}\,_a + \omega^I\,_{Ka} \omega^{KJ}\,_b - \omega^I\,_{Kb} \omega^{KJ}\,_a
- R^{IJ}\,_{ab} \right ) \approx 0.
\end{eqnarray}
Some of the differences with the theory in two dimensions are the following: (1) there are now velocities of the fields $T^I\,_{0a}$ and $R^{IJ}\,_{0a}$, (2) the new constraints $h^I$ and $H^{IJ}$ come from the fact that now $\phi_I$ and $\phi_{IJ}$ are one-forms. The $3+3+6+6+3+3=24$ constraints in Eq. (\ref{const2}) are first class and reducible because of the $3+3=6$ reducibility equations
\begin{eqnarray}
{\mathcal D}_a d^a\,_I - g_I + \frac12 \varepsilon_{ab} \Pi^{ab}\,_J H^J\,_I &=& 0, \nonumber\\
{\mathcal D}_a D^a\,_{IJ} - G_{IJ} + \frac12 \left ( d^a\,_I e_{Ja} - d^a\,_J e_{Ia} \right ) - \frac14 \varepsilon_{ab} \left ( \Pi^{ab}\,_I h_J - \Pi^{ab}\,_J h_I \right ) + \frac12
 \varepsilon_{ab} \left ( H^K\,_I \Pi^{ab}\,_{KJ} - H^K\,_J \Pi^{ab}\,_{KI} \right ) &=& 0. \qquad
\end{eqnarray}
Because of the fact that there are $6+6+3+3=18$ configuration variables, the number of local degrees of freedom is $\frac12 \left [ 2 (18) - 2(24-6) \right ]=0$ and so the action (\ref{action}) defines a topological field theory.



\begin{thebibliography}{99}
\bibitem{spinfoam}
A. Perez, Classical Quantum Gravity {\bf 20}, R43 (2003); D. Oriti, Rept. Prog. Phys. {\bf 64}, 1703 (2001); J.C. Baez, Lect. Notes Phys. {\bf 543}, 25 (2000); Classical Quantum Gravity {\bf 15}, 1827 (1998); J. Engle, R. Pereira, and C. Rovelli, Phys. Rev. Lett. {\bf 99} 161301 (2007); E.R. Livine and S. Speziale, Europhys. Lett. {\bf 81} (2008) 50004.
\bibitem{lqg}
T. Thiemann, {\it Modern Canonical Quantum General Relativity} (Cambridge University Press, Cambridge, England, 2007); C. Rovelli, {\it Quantum Gravity} (Cambridge University Press, Cambridge, 2004); A. Ashtekar and J. Lewandowski, Classical Quantum Gravity {\bf 21}, R53 (2004); A. Perez, ``Introduction to loop quantum gravity and spin foams,'' Proceedings of the International Conference on Fundamental Interactions, Domingos Martins, Brazil, (2004), arXiv:gr-qc/0409061.
\bibitem{cuesta}
V. Cuesta and M. Montesinos, Phys. Rev. D {\bf 76}, 104004 (2007).
\bibitem{jmp2006}
M. Mondrag\'on and M. Montesinos, J. Math. Phys. {\bf 47}, 022301 (2006).
\bibitem{iopmm}
M. Montesinos, J. Phys. Conf. Ser. {\bf 24}, 44 (2005).
\bibitem{cqg2006}
M. Montesinos, Class. Quantum Grav. {\bf 23}, 2267 (2006).
\bibitem{cqg2003}
M. Montesinos, Class. Quantum Grav. {\bf 20}, 3569 (2003).
\bibitem{dirac}
P.A.M. Dirac, {\it Lectures on Quantum Mechanics} (Belfer Graduate School of
Science, New York, 1964).
\bibitem{henneaux}
M. Henneaux and C. Teitelboim, {\it Quantization of Gauge Systems} (Princeton
University Press, Princeton, 1992).
\bibitem{cai95}
M.I. Caicedo, R. Gianvittorio, A. Restuccia, and J. Stephany, Phys. Lett.  B {\bf 354}, 292 (1995).
\bibitem{pibe}
M. Montesinos and A. Perez, Phys. Rev. D {\bf 77}, 104020 (2008).
\bibitem{baezp}
J.C. Baez and A. Perez, Adv. Theor. Math. Phys. {\bf 11}, 451 (2007).
\bibitem{winspibe}
W.J. Fairbairn and A. Perez, ``Extended matter coupled to BF theory'', arXiv:0709.4235v1 [gr-qc].
\bibitem{winston}
W.J. Fairbairn, R. Brasselet and A.Perez, ``Quantization of
string-like sources coupled to BF theory: physical scalar product
and spinfoam models'' (to appear).
\bibitem{karim}
K. Noui and A. Perez, Classical Quantum Gravity {\bf 22}, 4489 (2005);  Classical Quantum Gravity {\bf 22}, 1739 (2005); L. Freidel and D. Louapre, Classical Quantum Gravity {\bf 21}, 5685 (2004.
\bibitem{carlip}
S. Carlip, Nucl. Phys. B {\bf 324} (1989) 106; P. de Sousa Gerbert, Nucl. Phys. B {\bf 346}, 440 (1990).
\bibitem{cham}
A.H. Chamseddine and W. Wyler, Phys. lett. B {\bf 228}, 75 (1989).
\bibitem{cham2}
A.H. Chamseddine, Nucl. Phys. B {\bf 346}, 213 (1990).
\bibitem{cham3}
A.H. Chamseddine and D. Wyler, Nucl. Phys. B {\bf 340}, 595 (1990).
\bibitem{jmp1977}
J.F. Pleba\'nski, J. Math. Phys. {\bf 18}, 2511 (1977).
\bibitem{penrose}
M. P. Reisenberger, Nucl. Phys. B {\bf 457}, 643 (1995).
\bibitem{capo}
R. Capovilla, J. Dell, and T. Jacobson,  Class. Quantum Grav. {\bf 8}, 59
(1991).
\bibitem{cqg1999}
M.P. Reisenberger, Class. Quantum Grav. {\bf 16}, 1357 (1999).
\bibitem{cqg1999b}
R. De Pietri and L. Freidel, Class. Quantum Grav. {\bf 16}, 2187 (1999).
\bibitem{rob}
D.C. Robinson, J. Math. Phys. {\bf 36}, 3733 (1995).
\bibitem{cqgl2001}
R. Capovilla, M. Montesinos, V.A. Prieto, and E. Rojas, Class. Quantum Grav.
{\bf 18}, L49 (2001).
\bibitem{fkp}
L. Freidel, K. Krasnov, and R. Puzio, Adv. Theor. Math. Phys. {\bf 3}, 1289 (1999).
\end{thebibliography}
\end{document}